\documentclass[graybox]{svmult}

\usepackage{mathptmx}       
\usepackage{helvet}         
\usepackage{courier}        
\usepackage{type1cm}        
\usepackage{makeidx}         
\usepackage{graphicx}        
\usepackage{multicol}        
\usepackage[bottom]{footmisc}

\makeindex             

\begin{document}

\title*{Preliminary results on a Virtual Observatory search for companions
to Luyten stars}
\titlerunning{Preliminary results on a Virtual Observatory search for companions
to Luyten stars} 
\author{J.~A. Caballero, F.~X. Miret, J. Genebriera, T. Tobal, J. Cairol, D.
Montes}
\institute{J.~A. Caballero and D. Montes \at Departamento de Astrof\'{\i}sica y
Ciencias de la Atm\'osfera, Facultad de F\'{\i}sica, Universidad Complutense de
Madrid, E-28040 Madrid, Spain,
\email{caballero@astrax.fis.ucm.es}
\and
F. X. Miret, T. Tobal and J. Cairol \at Observatori Astron\`omic del Garraf,
Barcelona, Catalunya, Spain 
\and
J. Genebriera \at Observatorio de Tacande, El~Paso, La~Palma, Islas Canarias,
Spain}
%
\maketitle

\abstract*{The Aladin sky atlas of the Virtual Observatory has shown to be a powerful and
easy-handling tool for the discovery, confirmation, and characterisation of
high proper-motion, multiple stellar systems of large separation in the solar
vicinity.
Some of these systems have very low mass components (at the star/brown dwarf
boundary) and are amongst the least bound systems found to date. 
With projected physical separations of up to tens of thousands astronomical
units, these systems represent a challenge for theoretical scenarios of
formation of very low-mass stars and brown dwarfs.
Here we show preliminary results of a novel ``virtual'' search of binary systems
and companions to Luyten stars with proper motions between 0.5 and
1.0\,arcsec\,a$^{-1}$.}

\abstract{The Aladin sky atlas of the Virtual Observatory has shown to be a powerful and
easy-handling tool for the discovery, confirmation, and characterisation of
high proper-motion, multiple stellar systems of large separation in the solar
vicinity.
Some of these systems have very low mass components (at the star/brown dwarf
boundary) and are amongst the least bound systems found to date. 
With projected physical separations of up to tens of thousands astronomical
units, these systems represent a challenge for theoretical scenarios of
formation of very low-mass stars and brown dwarfs.
Here we show preliminary results of a novel ``virtual'' search of binary systems
and companions to Luyten stars with proper motions between 0.5 and
1.0\,arcsec\,a$^{-1}$.}

\section{Introduction}
\label{sec:1}

On the one hand, Caballero (2007b) showed that a minimum of 5.0$\pm$1.8\,\% of
the solar-neighborhood dwarfs with spectral types later than M5.0 (i.e. with
very low masses, $M \le$ 0.1\,$M_\odot$) are part of multiple systems of wide
separation ($r >$ 100\,AU).  
On the other hand, most of the objects in the {New Luyten Two Tenths} (NLTT)
catalogue (Luyten 1979) are nearby late-type dwarfs ($d \le$ 30\,pc) and
solar-like stars with very high proper motions.  
As a matter of fact, the primary stars of the two multiple systems whose common
proper motions were first measured by Caballero (2007b) are two stars in the
NLTT catalogue, {Koenigstuhl~2A} (NLTT~13422, early M) and {Koenigstuhl~3A}
(NLTT~57119, F8V). 
Koenigstuhl~2AB is one of the lowest mass, widest binaries yet found;
Koenigstuhl~3ABC is the widest (hierarchical, triple) system with an L-type
component ($\rho = 7.530\pm0.007$\,arcmin, $r = 11\,900\pm300$\,AU).
{Koenigstuhl~3C} (also known as 2MASS~J23310161--0406193~B), has a mass at the
stellar-substellar boundary.

The primary star in the binary system Koenigstuhl~1AB, discovered by
Caballero (2007a) using an identical search method, although it is too faint for
appearing in the NLTT catalogue, had been also previously tabulated in a
catalogue of high proper motion objects (Liverpool-Edinburgh survey; Pokorny
et~al. 2003). 
Koenigstuhl~1AB is the second widest system with $M_{\rm A} + M_{\rm B} <$
0.2\,$M_\odot$ ($\rho$ = 1.2956$\pm$0.0012\,arcmin, $r$ = 1800$\pm$170\,AU); the
widest system, 2M0126--50~AB, was afterwards discovered by Artigau
et~al.~(2007). 
The importance of systems with very low binding energies for theoretical
scenarios of formation of substellar objects have been already discussed in
detail in, e.g., Phan-Bao et~al. (2005), Bill\`eres et~al. (2005), Burgasser
et~al. (2007), and Artigau et~al. (2007). 
They can also be used as a ``test particle'' to investigate the gravitational
potential of the Galaxy (Caballero~2007a).

Given the relatively small amount of field late-type stars and brown dwarfs
studied by Caballero (2007b; he investigated only 173 targets), it is natural to
design a new massive survey to explore a larger quantity of dwarfs.
Since the larger the proper motion of a system is, the easier the confirmation
of common proper motion in a defined time base line is, one should look for
companion candidates to the {\em fastest} stars.
The vast majority of them were tabulated by, as the reader may guess, in the
NLTT catalogue, which tabulates several tens of thousand stars with proper
motions larger than {\em two tenths} arcseconds per annum ($\mu >$
0.2\,arcsec\,a$^{-1}$).
Salim \& Gould (2003) improved the astrometry of about 36\,000 of these stars
from (optical) USNO-A\footnote{USNO-A: U.S. Naval Observatory catalogue of
astrometric standards (Monet et~al. 1998).} and (near-infrared)
2MASS\footnote{2MASS: Two Micron All Sky Survey (Skrutskie et~al. 2006).} data.
Our results will complement other searches for Luyten binaries with high proper 
motions (e.g. Oswalt et~al. 1988; Ryan 1992; Allen et~al. 2000; L\'epine et~al.
2002; Salim \& Gould 2003; Chanam\'e \& Gould 2004; Levine 2005). 

\section{Analysis} 
\label{sec:2}

We have used the survey method described in Caballero (2007a, 2007b), but
with a slight difference:
instead of using the SuperCOSMOS Science Archive (Hambly et~al. 2001), we have
employed the USNO-B1\footnote{USNO-B1: U.S. Naval Observatory catalogue of
astrometric standards (Monet et~al. 2003).} and 2MASS catalogues and, in an
intensive and extensive way, the Virtual Observatory tool Aladin (Bonnarel
et~al. 2000).
Although the astrometric precision and depth of USNO-B1 are worst than those
of SuperCOSMOS, we have still some fundamental adavantages:
USNO-B1 covers the whole sky (not only the Southern Hemisphere, as SuperCOSMOS
does), a mathematical environment for plotting and selecting (e.g. Matlab, IDL)
is unnecessary, and, especially, Aladin allows to carry out an interactive
search {\em very quickly} (literally, in ``a few mouse clicks'').
An experienced individual with a home internet bandwidth can identify a
common proper-motion companion candidate in less than two minutes.
The basic scheme of the analysis is as follow:

\begin{figure*}[t]
\sidecaption
\includegraphics[width=0.49\textwidth]{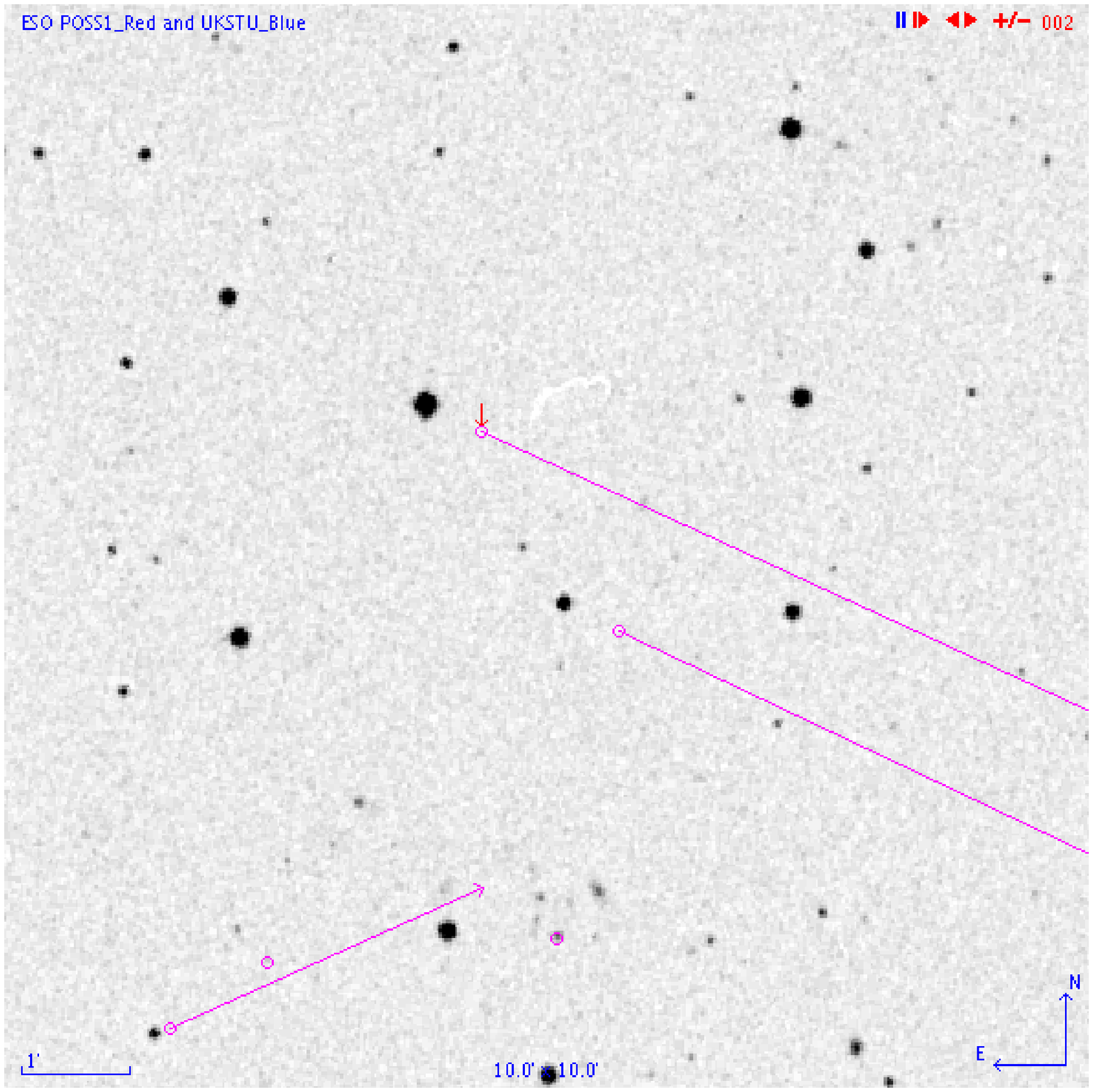}
\includegraphics[width=0.49\textwidth]{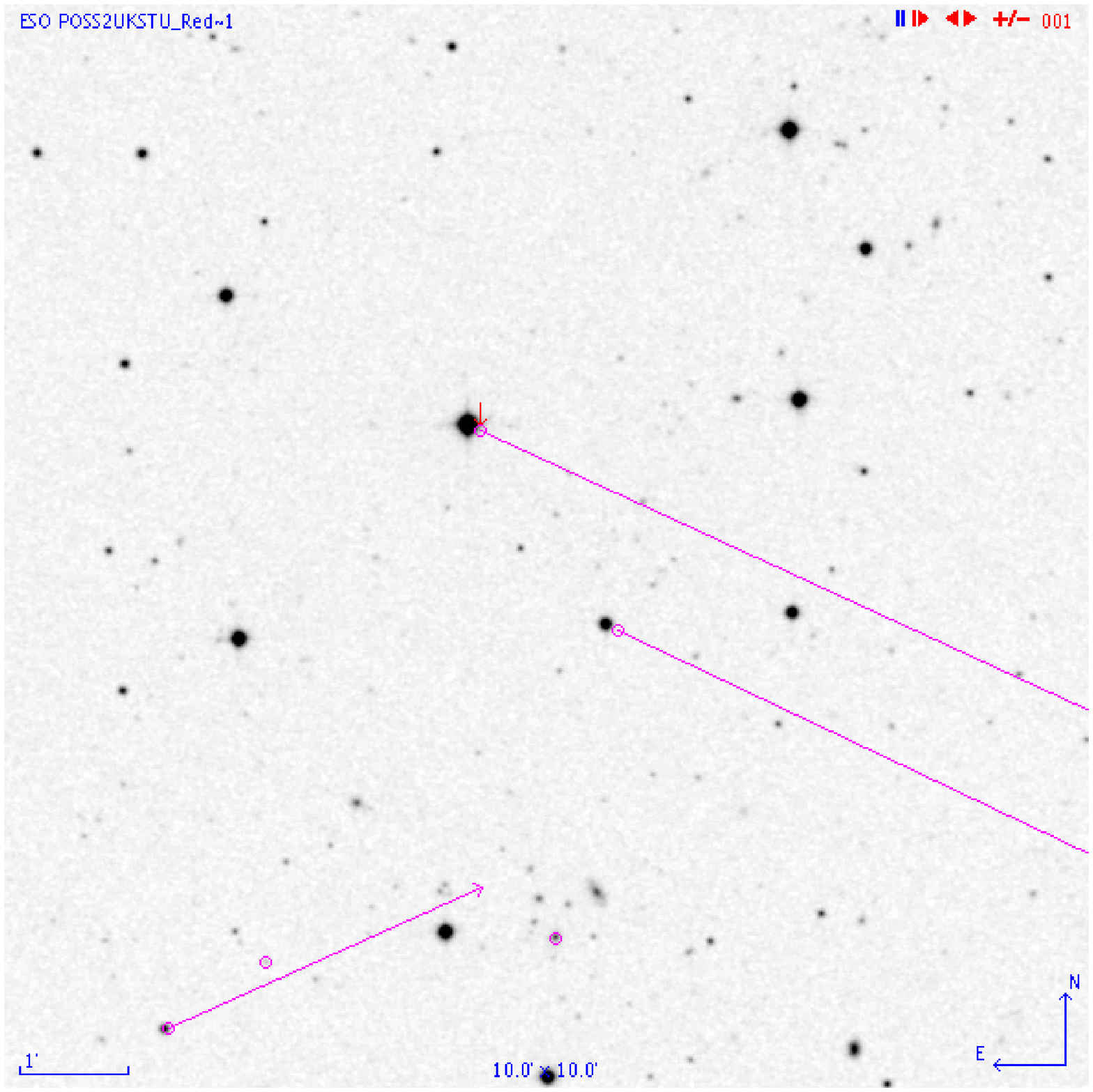}
\caption{Colour-inverted, 10$\times$10\,arcmin$^2$-size, ESO POSS1 Red
(J1950.379, {\em left}) and POSS2 UKST Red (J1989.176, {\em right}) images
centred on the known common proper motion system G~200--15 (NLTT~36000, sd:K4)
and vB~7 (NLTT~35991, M7V).
Angular separation and position angle of the system are $\rho \approx$
2.21\,arcmin and $\theta \approx$ 214\,deg.
It was identified in the proper motion diagram centred on the primary.
North is up and East is left.
The high proper motion star to the Southeast of the system is LP~174--32, which
has a different spatial velocity.}  
\label{fig_garraf052}       
\end{figure*}

\begin{enumerate} 
\item choose the Luyten star to which companion candidates will be searched for.
We have selected for investigation the 1947 stars (and white dwarfs) in Salim \&
Gould (2003) with proper motions in the interval 0.5--1.0\,arcsec\,a$^{-1}$,
\item load a red 30$\times$30\,arcmin$^2$ Digitized Sky Survey image centred on
the star in the view window of Aladin,
\item load all the targets in the Simbad astronomical database and in the 2MASS
and USNO-B1 surveys in a 15\,arcmin-radius circle centred on the star,
\item cross-match 2MASS and USNO-B1 (in this order) with the default
cross-matching radius of 4\,arcsec,
\item identify common proper motion pairs with 2MASS+USNO-B1 data:
\begin{enumerate} 
	\item plot a proper motion diagram ($\mu_\delta$ vs. $\mu_\alpha
	\cos{\delta}$) of the cross-matched sources with VOplot,
	\item mark the position of the Luyten star in the diagram,
	\item study if there are objects in the area with proper motions similar
	to those of our star,
	\item if there are, check that they are not faint optical/near-infrared
	sources with incorrect USNO-B1 proper motions,
\end{enumerate} 
\item identify common proper motion pairs with Simbad data:
\begin{enumerate} 
	\item activate the ``proper motion'' dedicated filter,
	\item study if there are objects in the area with proper motions similar
	to those of our star,
\end{enumerate} 
\item and, finally, if a common proper motion pair candidate exists,
characterise it in detail.  
Given their brighness and wide angular separations, most of the pair candidates
can be easily followed-up with small telescopes (e.g. 40\,cm) at amateur
observatories.
\end{enumerate} 

\begin{figure*}[t]
\sidecaption
\includegraphics[width=0.49\textwidth]{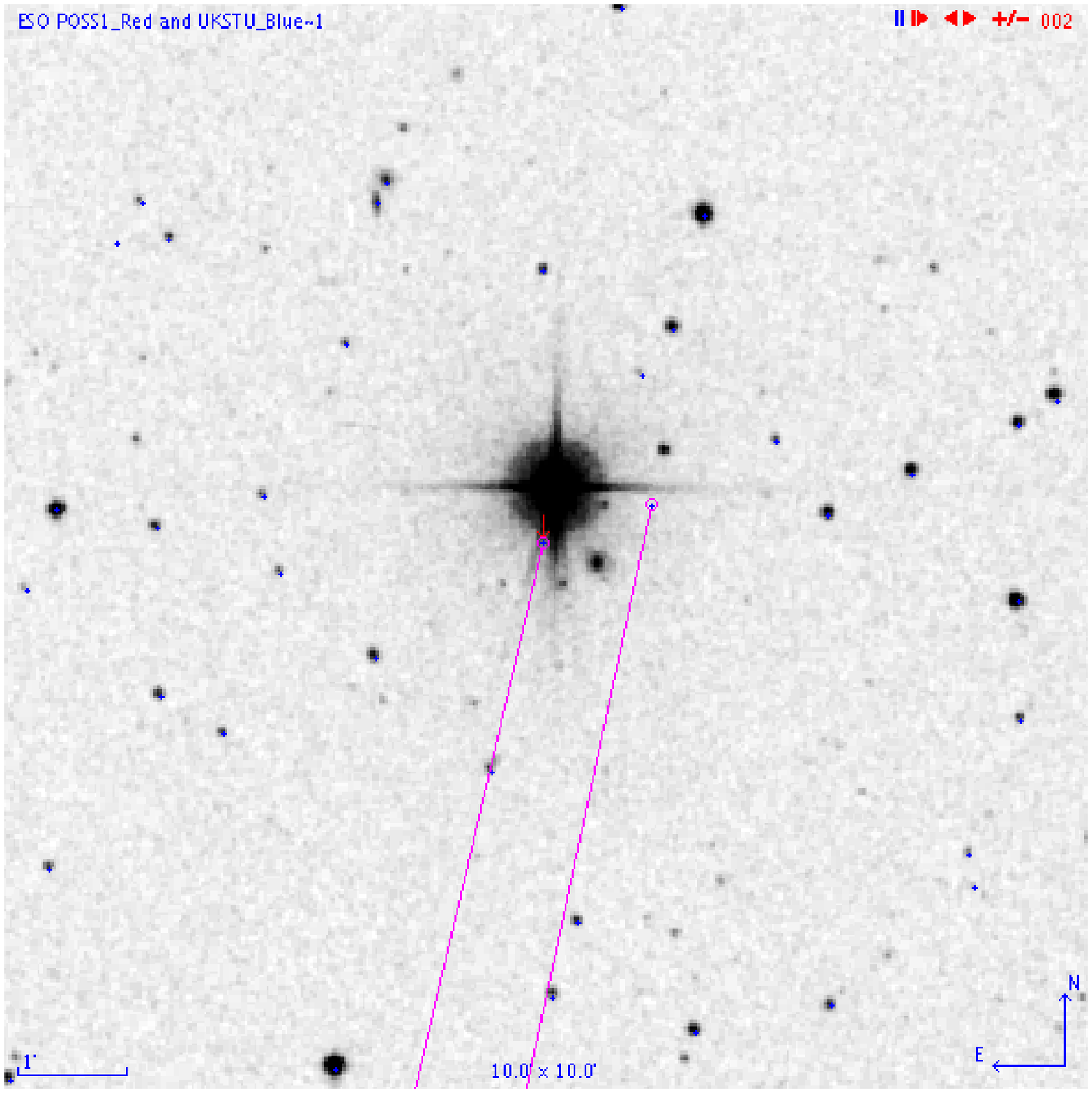}
\includegraphics[width=0.49\textwidth]{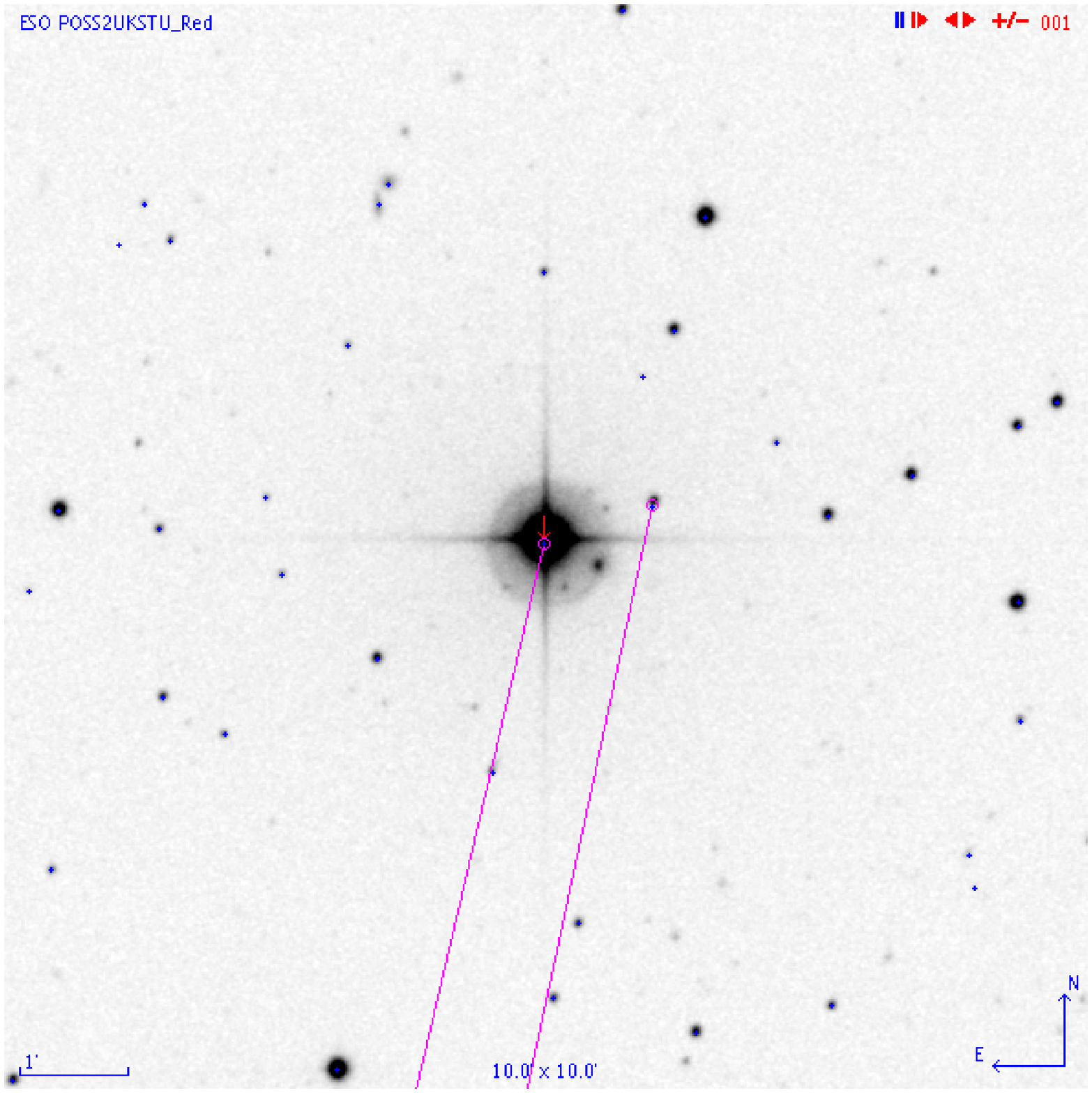}
\caption{Colour-inverted, 10$\times$10\,arcmin$^2$-size, ESO POSS1 Red
(J1950.218, {\em left}) and POSS2 UKST Red (J1996.389, {\em right}) images
centred on the {\em new} common proper motion system HD~126512 (NLTT~37293, F9V,
$d_{\rm HIP}$ = 46.9$\pm$1.8\,pc) and LSPM~J1425+2035W (a poorly known high
proper motion star catalogued only by L\'epine \& Shara 2005). 
Angular separation and position angle of the system are $\rho \approx$
1.05\,arcmin and $\theta \approx$ 290\,deg.
It was identified with the Simbad ``proper motion'' filter.
The secondary has the optical/near-infrared magnitudes and colours of a
mid/late-K or early-M dwarf at $d \sim$ 50\,pc.
North is up and East is left.} 
\label{fig_garraf124}       
\end{figure*}

\noindent We will give details on the steps of the analysis (including 
selection of the input list of Luyten stars, data loading and
cross-matching, proper motion diagram and candidate selection, and 
astrometric follow-up) in a forthcoming paper.

\section{Preliminary results} 
\label{sec:3}

\begin{figure*}[t]
\sidecaption
\includegraphics[width=0.49\textwidth]{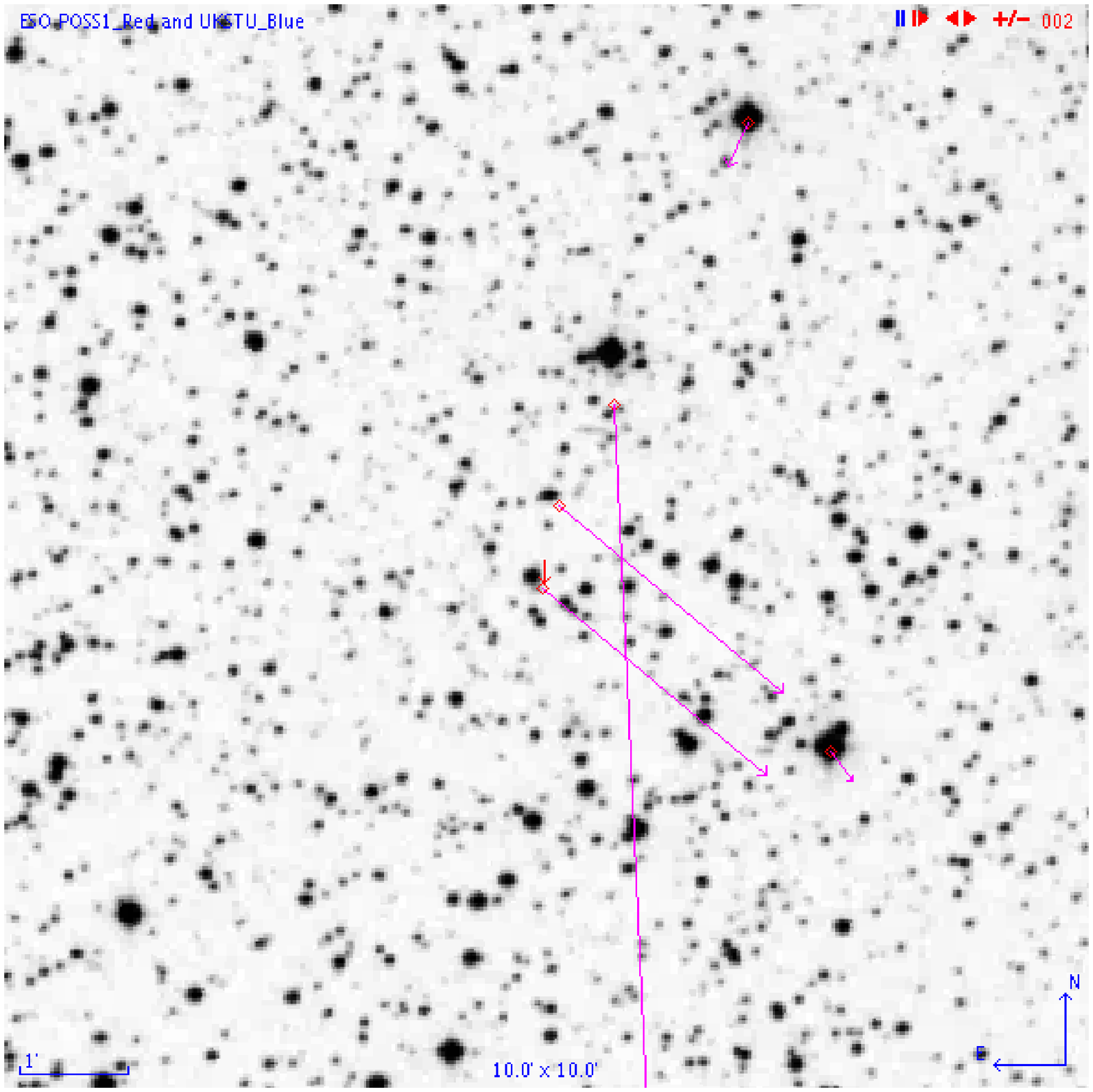}
\includegraphics[width=0.49\textwidth]{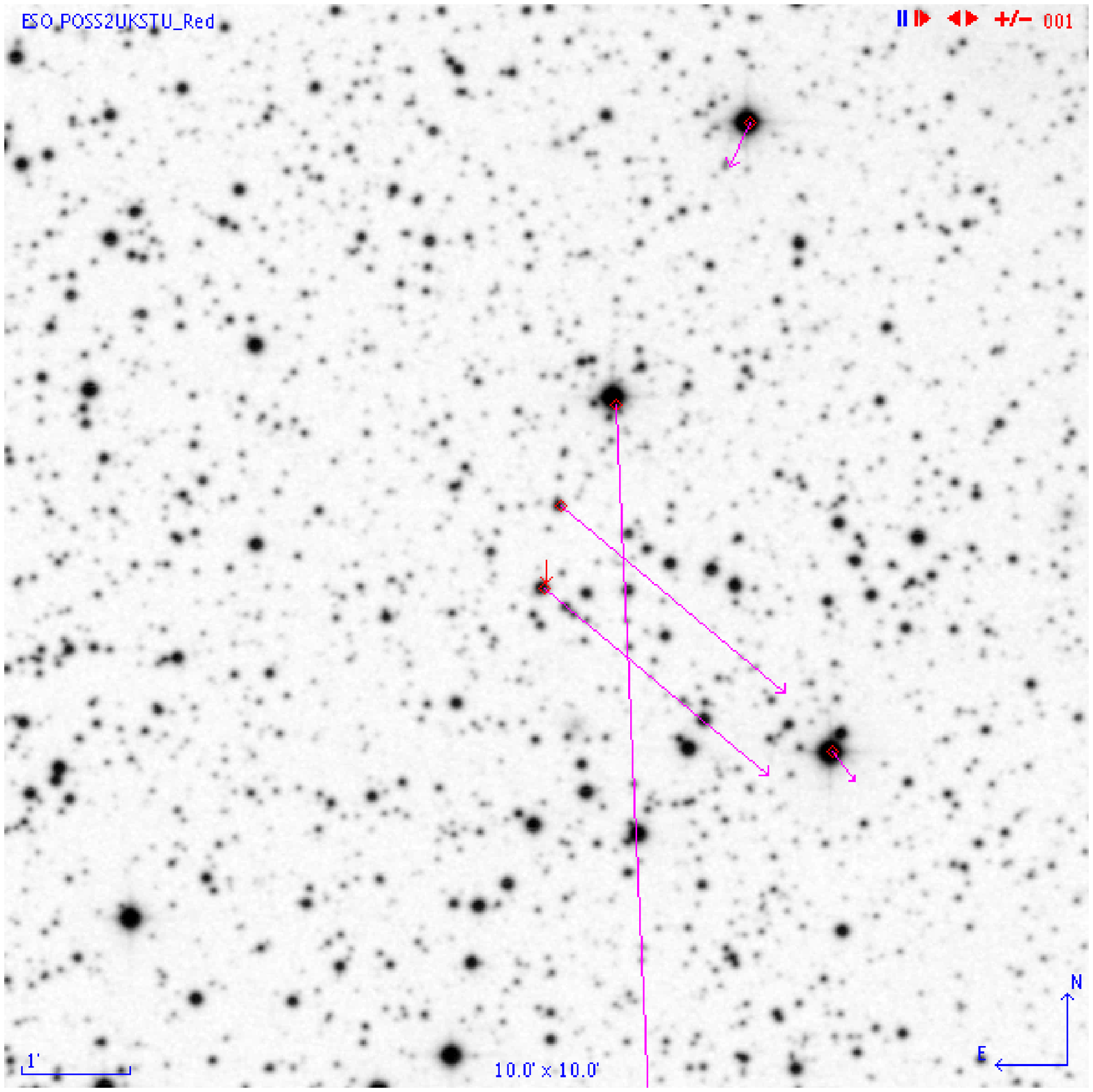}
\caption{Colour-inverted, 10$\times$10\,arcmin$^2$-size, ESO POSS1 Red
(J1952.542, {\em left}) and POSS2 UKST Red (J1992.664, {\em right}) images
centred on the common proper motion system G~125--15 (NLTT~47904, M4.5,
pulsation variable, X-ray) and G~125--14 (NLTT~47903, M5--6:). 
Angular separation and position angle of the system are $\rho =$
0.766$\pm$0.005\,arcmin and $\theta =$ 347.34$\pm$0.16\,deg.
The absolute value of its common proper motion, $\mu$ = 0.16\,arcsec\,a$^{-1}$,
prevented the system to be listed in the input catalogue. 
The binary was serendipitously identified in the proper motion diagram centred
on the high proper motion star BD+35~3659 (NLTT~47900, F1, at about 1\,arcmin to
the North of the pair). 
North is up and East is~left.} 
\label{fig_garrafXXX}       
\end{figure*}

We give some preliminary results on our Virtual Observatory search for
companions to Luyten stars. 
Of the 1947 stars in the input catalogue with proper motions in the interval
0.5--1.0\,arcsec\,a$^{-1}$, we have identified 101 of such stars in 52 systems
identified in common proper motion diagrams.
All the systems are pairs of Luyten stars except for a hierarchical triple
(GJ~421~ABC) and four binaries with secondaries not present in Salim \& Gould
(2003).
From the unfinished analysis of systems pre-selected with the Simbad ``proper
motion'' filter, we have identified 46 Luyten stars in 24 binaries.
The primaries with no Luyten star as a secondary are GJ~383 and HD~126512 (see
below).
The analysis of another pre-selected $\sim$30 Luyten stars is on-going. 

The identified systems are subject to an astrometric characterisation after
classyfing them into Luyten multiple systems with both components in the
{\em Hipparcos} catalogue, with only one component in {\em Hipparcos}, and with
no component in {\em Hipparcos}.
Most of the systems in the third group have been poorly characterised or not
characterised at all.
Besides, we have serendipitously identified 10 additional binaries with proper
motions less than 0.5\,arcsec\,a$^{-1}$ in the survey areas of the Luyten stars in
the input catalogue.

We show in Figs.~\ref{fig_garraf052}, \ref{fig_garraf124} and
\ref{fig_garrafXXX} three examples of binary systems identified in proper
motion diagrams and with the Simbad ``proper motion'' filter, including an
interesting ``serendipitous'' pair.
The system G~125--15 + G~125--14 (Fig.~\ref{fig_garrafXXX}) is a bright,
young, very wide, low-mass binary in the solar neighbourhood (Caballero et~al.,
in~prep.).
At only $d \sim$ 17\,pc, the most probable projected physical separation and
individual masses are $\Delta \sim$ 800\,AU, $M_A \sim$ 0.2\,$M_\odot$ and $M_B
\sim$ 0.1\,$M_\odot$, which correspond to one of the lowest binding energies
measured in a binary. 
It is being subject of a detailed multi-epoch, multi-band, astrometric,
photometric and spectroscopic follow-up because Daemgen et~al. (2007) proposed
the primary to be 300--500\,Ma old (i.e. individual masses could be lower).

\begin{acknowledgement}
JAC thanks Bertrand Goldman for first suggesting the extension of the
Koenigstuhl survey to Luyten stars.
Partial financial support was provided by the Universidad Complutense de Madrid,
the Spanish Virtual Observatory, the Spanish Ministerio Educaci\'on y Ciencia,
and the European Social Fund under grants AyA2005--02750, AyA2005--04286,
AyA2005--24102--E, and AyA2007--67458 of the Programa Nacional de
Astronom\'{\i}a y Astrof\'{\i}sica and by the Comunidad Aut\'onoma de Madrid
under PRICIT project S--0505/ESP--0237 (AstroCAM).  
\end{acknowledgement}

\end{document}